# DSA SECURITY ENHANCEMENT THROUGH EFFICIENT NONCE GENERATION

Akash Nag[*1], Sunil Karforma[2]

[*1]Department of Computer Science, The University of Burdwan, Burdwan, West Bengal, India
nag.akash.cs@gmail.com[1]
[2]Department of Computer Science, The University of Burdwan, Burdwan, West Bengal, India
sunilkarforma@yahoo.com[2]

*Abstract*: The Digital Signature Algorithm (DSA) has become the *de facto* standard for authentication of transacting entities since its inception as a standard by NIST. An integral part of the signing process in DSA is the generation of a random number called a nonce or an ephemeral key. If sufficient caution is not taken while generating the nonce, it can lead to the discovery of the private-key paving the way for critical security violations further on. The standard algorithms for generation of the nonce as specified by NIST, as well as the widely implemented random number generators, fail to serve as true random sources, thus leaving the DSA algorithm open to attack, resulting in possible signature forgery in electronic transactions, by potential attackers. Furthermore, the user can select the nonce arbitrarily, which leads to a subliminal channel being present to exchange messages through each signature, which may be intolerable for security reasons. In this paper, we have improved the security of the DSA algorithm by proposing an efficient nonce-generation process, which ensures that the generated nonce is sufficiently random as well as unique for each generated signature, thereby securing the signing process. Furthermore, our algorithm also ensures that there are no subliminal channels present in DSA.

## 1. INTRODUCTION

A digital signature scheme is a process by which a receiver can verify the authenticity of the origin of a message, as well as be assured of its integrity simultaneously. The scheme is based on the public-key cryptographic system [1] and requires that the signer maintain two keys – one private (or secret-key), and another, a public-key; the latter assumed to be safely maintained by a Trusted Centre (TC), and is available to anyone. The private-key is strictly and confidentially maintained by its owner. A message-digest of the file to be signed, is computed first, followed by applying the signing algorithm on this digest using the signer's private key. Optionally, instead of the digest, the entire message can be used as well, however, using the digest instead of the message is computationally more efficient for signing purposes. The original file, the generated signature, and optionally the signer's public-key, are sent to the receiver, who re-computes the digest and verifies the signature using the signer's public-key. If the verification succeeds, the message is authentic (i.e. it is indeed the claimant who has sent the message), and it has not been altered in transit. The security of this scheme lies in the computational difficulty in solving the discrete logarithm problem [18], which would otherwise lead to the discovery of the signer's private key from the corresponding public-key.

The signing process requires the generation of an integer, called the nonce or ephemeral key, which must be generated randomly or pseudo-randomly, and must be unique for all messages originating from that sender. If an attacker ever recovers the nonce that has been used to sign the message, perhaps by exploiting some properties of the pseudorandom number generator used to generate it, he/she can recover the signer's private-key. Moreover, if the signer uses the same nonce to sign two messages, the attacker can recover the private-key even if he/she is unaware of the actual value of the nonce used [13] [14].

The Digital Signature Standard (DSS) [2] suggests two ways for generating the nonce, but both suffer from being vulnerable to lattice attacks as shown by various papers [5] [7] [8] [9], due to the use of a Linear Congruential Random Bit Generator (RBG), which suffers from the flaw that they are predictable once the seed used to initialize it is revealed. RBGs are of the form: $f(n)=(a+b\times n)(\bmod m)$, where a, b, and m are parameters to the function. Seemingly random numbers are returned by the function because the numbers form a part of a very large sequence that is unavailable for instant comparison. However, the values for these parameters can be estimated if a large number of nonces are mistakenly revealed.

The nonce, when instead of being generated randomly, is rather chosen arbitrarily by the signer, reveals another flaw in the DSA scheme. There exists a subliminal channel [17] in DSA, through which the signer and the verifier can communicate secretly by exchanging messages through each signature. Since however the nonces must be random, and messages, by nature, are non-random – the signer must encrypt the messages, preferably using a one-time pad, and then use this encrypted random message as the nonce. It should be clear that the maximal length of the message should be that of the size of the nonce. The receiver, on the other hand should be aware of the signer's private-key to obtain this nonce, and hence, the message.

Our proposed scheme generates nonces securely, and is not based on Linear Congruential number generators. Instead, it uses Cryptographically Secure Pseudo-Random Number Generators (CSPRNGs) as well as additional masking processes, to make the nonce generation process foreseeably secure. Section 2 of this paper recapitulates on the Digital Signature Algorithm as laid down in the NIST's standard, followed by an overview of the attacks on DSA in Section 3, where several lattice as well as non-lattice attacks are mentioned, specifically targeted at the nonce generation process of DSA. Section 4.1 presents the proposed algorithm for secure nonce generation, and Sections 4.2 – 4.10 explain the key steps of the algorithm. Finally in Section 5, we conclude with our results and an informal security analysis of our algorithm, stating the advantages of implementing it in any signature generation scheme based on DSA.

## 2. DIGITAL SIGNATURE ALGORITHM

The National Institute of Standards and Technology (NIST) of the U.S. Govt. published the Digital Signature Standard (DSS) under FIPS PUB 186-3 [2], which proposed an algorithm, known as Digital Signature Algorithm (DSA), for generation and verification of digital signatures. A brief review of the algorithm is given below. In DSA, a Trusted Center (TC) is responsible for generating and publishing system parameters for public usage.

### 2.1 Parameter Sizes
(L, N): it is a bit-length pair for generation of system parameters. For the purposes of our algorithm: L=2048 and N=224.

### 2.2 System Parameters
p: A prime modulus, where $2^{L-1} < p < 2^L$
q: A prime divisor of (p-1), where $2^{N-1} < q < 2^N$
g: $h^{(p-1)/q}$ (mod p), where h is a random integer with $1 < h < p$

### 2.3 Private and Public Keys
x: each user selects an integer x as his private-key, with $1 < x < q$
y: corresponding to x, the user computes $y = g^x$ (mod p)

### 2.4 Signature Generation
The signer computes the following values corresponding to a message m to be signed:

k: a secret number that is unique to each message; k is an integer, which is generated randomly or pseudo-randomly, in the range [1, q-1]. (From here on in this document, k is referred to as the "nonce")
$r = (g^k \pmod p) \pmod q$
$s = (k^{-1} \cdot (H(m) + x \cdot r)) \pmod q$ where, H(m) is a one-way keyless hash function

The pair (r, s) constitutes the digital signature of the message m, signed by the user with public-key y and private-key x.

### 2.5 Signature Verification
To verify the signature pair (r, s), the recipient computes:

$w = s^{-1} \pmod q$
$u_1 = (w \times H(m)) \pmod q$
$u_2 = (w \times r) \pmod q$
$r' = (g^{u_1} \times y^{u_2} \pmod p) \pmod q$

If r=r' the signature is authentic, otherwise not.

## 3. ATTACKS ON DSA

### 3.1 Computational Basis of DSA Security

The security of DSA is based on the computational difficulty in solving the discrete logarithm problem in prime fields and its subgroups, and it can be proved under the random oracle model [3], which assumes that the hash function behaves like a random oracle, i.e. its values are independent and uniformly distributed. It is cautioned in the standard that if the nonce (k) is disclosed, the secret-key can be easily recovered.

### 3.2 The Insecurity of Standard RBG
Standard Random-Bit Generators (RBGs) are typically used to generate random bits in software/hardware using a Linear Congruential Generator (LCG). One of the most commonly used is the Knuth's LCG [15]. It was shown by Belarre et al. [4] that the secret-key of DSA can be recovered if the nonce is generated by Knuth's linear congruential generator with known parameters. This attack is provable and relies on Babai's approximation algorithm [5], based on the LLL algorithm [6]. Howgrave-Graham and Smart [7] showed that, even if the nonce is known only partially, i.e. only some of its bits are revealed, for a reasonable number of signatures, a number of heuristic attacks are possible to recover the secret-key. Finally, Nguyen and Shparlinski [8] presented a provable polynomial-time attack against DSA when the nonces are partially known, under two assumptions: the size of q should not be too small compared to p, and the probability of collisions for the hash function H should not be too large compared to 1/q. Under these conditions, if for a certain number of random messages $\mu \in M$ and random nonces $k \in [1, q-1]$, about $\log^{(1/2)} q$ least significant bits of k are known, then in polynomial time, one can recover the signer's secret-key x.

## 4. SECURE NONCE GENERATION

### 4.1 The Proposed Algorithm

Our algorithm proposes to augment the existing signature algorithm with a secure nonce-generation process in place of the existing schemes recommended by DSS. Our algorithm (Algorithm I) uses a cryptographically secure pseudorandom number generator (CSPRNG) instead of the simple RBG with additional processing to improve security.

A brief outline of our algorithm (Algorithm I) is as follows:
a) The algorithm takes two parameters as input; first is the message or data to be signed (the first 2048 bytes of it, if the message is larger than that), stored in the byte-vector M, and the second is the signer's 224-bit private-key. The algorithm begins with padding and dividing the message into 512-bit blocks, while the 224-bit key goes through two rounds of key-expansions to generate a 512-bit key suitable for use in our algorithm. The initial key-expansion (Section 4.2) expands it from 224 bits to 256-bits. This is known as the Partially Expanded Key (PEK). The second round of expansion is the Rijndael expansion (Algorithm II), (which is the standard Rijndael key-expansion [10] routine performed twice), expanding it further to 512-bits (Section 4.4). This is now referred to as the Fully Expanded Key (FEK).
b) The algorithm then iterates 32 times, once for each of the 512-bit blocks. In each round, the following steps are performed:
  i. A different 512-bit round-key is generated in each round using a CSPRNG.
  ii. The 512-bit message-block for this round undergoes a block-substitution (Section 4.6, Algorithm III) using the PEK as the key, to obtain the Substituted-Message-Block (SMB). The substitution is achieved in a Feistel network of 8 rounds, and each round consists of the following:
    a) A different 256-bit round-key is generated (Algorithm IV) for each round using the PEK as input.

b) The message-block is divided into two halves, of 32-bytes each. A copy of the left-half and the round-key are then encrypted (using Algorithm V).
c) The right-half is XOR-ed with the encrypted block, and finally, the data for the next round is formed by concatenation of the original left-half and the currently modified right-half.
iii. The SMB and the Round-Key is passed through the Shabal Key-Permutation Schedule [11] for a permutation operation (Section 4.7), to obtain a 896-bit output.
iv. The 896-bit permuted block undergoes a compression (Section 4.8, Figure 1) followed by a substitution (Section 4.9) using the non-invertible S-Box Z (see Table I) to obtain a 512-bit output, known as the Compressed-Permuted Message-Block (CPMB).
v. The FEK is modified to store the bitwise XOR of the CPMB and the current value of the FEK.
c) The modified FEK is hashed using HMAC SHA-512, using the PEK as the key, to obtain the digest (Section 4.8).
d) The digest obtained in the previous step is compressed (Section 4.8, Algorithm VI), from 512-bits to 224-bits, to finally generate the nonce.

A detailed description of the algorithm steps are outlined below.

Algorithm I. Nonce Generation

**Input:** The message to be signed ($M$), the signer's 224-bit private-key ($S$)
**Output:** The 224-bit *nonce* or ephemeral key

**begin**
    $X$ = ExpandKey($S$);
    $B[1..n]$=Pad($M$);
    $F$=RijndaelExpand($X, A$);        // Algorithm II
    **for** $i$=1 **to** $n$ **do**
    **begin**
        $R$=CSPRNG_Random( );
        $T$=BlockSubstitute($X, B[i]$);    // Algorithm III
        $C$=CompSub(ShabalPermute($T, R$));
        $F$=$F$ **xor** $C$;
    **end**
    $K$=HMAC_SHA512($F, X$);
    **return** Compress($K$);        // Algorithm VI
**end**

### 4.2 Initial Key Expansion

Our algorithm requires a 256-bit key, but the signer's private-key is only of 224-bits. This calls for a key expansion, which randomly selects 4-bytes out of the 28-bytes making up the key. These 4-bytes are then appended to get a 32-byte key to be used for the subsequent steps.

### 4.3 Message Padding and Division into Blocks

The message or data is first null-padded to make it a multiple of 512-bits. It is then divided into equal blocks of 512-bits each. Only the first 2048 bytes of the message is considered. The value of n is set to the number of blocks so obtained, where $1 \leq n \leq 32$.

### 4.4 Final Key Expansion (Rijndael Expansion)

The expanded secret-key (256-bit) and the constant A (see Table I) are passed to the 2nd key expansion routine for a further round of expansion, from 256-bits to 512-bits using the Rijndael [10] key-expansion algorithm (Algorithm II). The key-expansion is achieved by dividing the input into 2 sets of 128-bits each, and then expanding each set to 256 bits. The state array is a set of two 4x4 byte-matrix, and the key is copied to it in column-major fashion, and subsequently read out row-wise as 32-byte blocks. These are then XOR-ed with the message-block. The byte-substitution made by the subroutine G (from the standard Rijndael subroutine) uses the Rijndael S-Box to perform the substitution. The value returned by this subroutine is XOR-ed with the modified data-block to initialize an extended data-block. This is then used as the seed in the subsequent iterative rounds, in each step of which, an extended data-block is generated on the basis of the previous extended-block and its corresponding modified-data-block. The final 512-bit output is generated by alternate concatenations of one data-block and one extended-data-block.

Algorithm II. Final Key Expansion (*Rijndael* Expansion)

**Input:** The 256-bit expanded-private-key ($K$) and 512-bit data-block ($M$)
**Output:** Expanded 512-bit private-key

**begin**
    declare $W_e$[0..1][0..3];
    $S$=CopyKeyIntoStateArray($K$);
    $W$[0..1][0..3]=GetWordsFromStateArray($S$);
    $W$=XorWithData($W,M$);
    **for** $j$=0 **to** 1 **do**
    **begin**
        $W_e[j][0]$=$W[j][0]$ **xor** G($W[j][3], j$+1);
        **for** $i$=1 **to** 3 **do**
        **begin**
            $W_e[j][i]$=$W_e[j][i$-1] **xor** $W[j][i]$;
        **end**
    **end**
    **return** Concatenate($W$[0], $W_e$[0], $W$[1], $W_e$[1]);
**end**

### 4.5 Secure Random Key Generation

The algorithm proceeds in n rounds, where n is the number of 512-bit blocks in the message. Each round requires a 512-bit random-key which is generated using a cryptographically secure pseudorandom generator and hence, is sufficiently unpredictable. Standard CSPRNG implementations are widely available in various programming language libraries.

### 4.6 Initial Block Substitution

In each round, the corresponding message-block is first combined with the key using a substitution algorithm (Algorithm III). The process consists of 8 rounds with each round using a different round-key, generated from the original key using a round-key generation algorithm. The round-key for a round is generated, using Algorithm IV, by circularly left-shifting the round-key of the previous round by 1 or 2 bits (depending on the particular round), and then XOR-ing it with the constant P (see Table I). The rounds in the block-substitution process implement a Feistel [16] cipher with block-size of 512-bits. At each step, the data is divided into two halves of 256-bits each. The left-half is the same as the right-half generated after the previous-round. The right-half is obtained by XOR-ing the previous left-half and the output of the round-cipher algorithm, which takes the previous-right-half and the round-key as parameters. The round-cipher algorithm (Algorithm V) performs a byte substitution using the Rijndael S-Box on the data and XORs it with the round-key to return the result.

## 4.7 Block Permutation (Shabal Permutation)

After block substitution, the data undergoes a keyed permutation using the Shabal [11] key permutation algorithm, and the round-key generated in the previous step. The permutation result consists of a pair of blocks of 384 and 512 bits respectively, which are then concatenated to get a 896-bit output. This output is then fed to the CompSub function (see Algorithm I), which consists of 2 steps: Block compression followed by a second block substitution.

Algorithm III. Initial Block Substitution

**Input:** 256-bit expanded-key (K), 512-bit message-block (M)
**Output:** 512-bit substituted message-block
**begin**
    roundKey=K;
    data=M;
    **for** i=1 **to** 8 **do**
        **begin**
        roundKey=GenerateRoundKey(roundKey, i);   // Algorithm IV
        left=data[32..63];
        temp=left;
        right=data[0..31];
        F=RoundCipher(temp, roundKey);        // Algorithm V
        **for** j=0 **to** 31 **do**
            right[j]=right[j] **xor** F[j];
        data=Concatenate(left, right);
    **end**
    **return** data;
**end**

## 4.8 Block Compression

The permuted output from the Shabal algorithm is then compressed using a 7:4 bit compression, proceeding from the most significant bit to the least significant in groups of 7 bits and replacing each group with a 4-bit nibble using XOR operations, as shown in Fig. 1, to get a 512-bit output.

## 4.9 Second Block Substitution

The 512-bit output from the block compression phase is fed into a substitution phase, which is different from the block substitution performed in Algorithm III. Here, each of the 64 bytes is passed through a 16x16 non-invertible S-Box Z (see Table I). The result of the substitution is finally XOR-ed with the cumulative cipher-block G in the main nonce generation algorithm.

## 4.10 Digest Generation and Compression

The result from the Shabal permutation is passed to the digest generation function, which is a keyed HMAC using the SHA-512 algorithm to generate message digests. The resulting 512-bit digest is further compressed (Algorithm VI) to finally generate the 224-bit nonce. The compression process consists of 3 steps. Firstly, each successive pair in the digest is XOR-ed together, resulting in a 256-bit digest. This is followed by the first 4 bytes being XOR-ed with the corresponding last 4 bytes, and then discarding the last 4 bytes, leaving a 224-bit digest. Finally, this is XOR-ed with the constant Q (see Table I), to generate the nonce, which can then be used in the DSA signature generation process.

Algorithm IV. Round Key Generation

**Input:** Previous-round's 256-bit key (Key), Round number (r)
**Output:** 256-bit round-key for this round
**begin**
    **if** r<3 **then** K=1 **else** K=2;
    **for** i=0 **to** 31 **do**
        Key[i]=( Key[i] <<< K ) **xor** P[i];
    **return** Key;
**end**

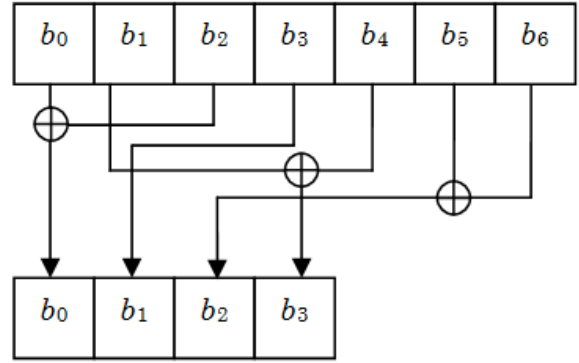

Figure 1. Block Compression.

Algorithm V. Round Cipher

**Input:** 256-bit data-block (data) and 265-bit round-key (key)
**Output:** 256-bit encrypted data-block

**begin**
    **declare** result[32];
    **for** i=0 **to** 31 **do**
        result[i]=key[i] **xor** Rijndael_Substitute(data[i]);
    **return** result;
**end**

Algorithm VI. Digest Compression

**Input:** 512-bit data (data)
**Output:** 224-bit nonce

**begin**
    **declare** result[28], temp[32];
    **for** i=0 **to** 31 **do**
        temp[i]=data[2*i] **xor** data[2*i+1];
    **for** i=0 **to** 3 **do**
        temp[i]=temp[i] **xor** temp[31-i];
    **for** i=0 **to** 28 **do**
        result[i]=temp[i] **xor** Q[i];
    **return** result;
**end**

## 5. RESULTS AND ANALYSIS

### 5.1 Results

Our proposed algorithm was implemented using Java [20]. The constants used in our algorithm are tabulated in Table I [19]. The running time of our algorithm is longer than the simple pseudorandom nonce generation process recommended in DSS. However, our algorithm is more secure, and is still fast, with a constant time complexity of O(1) since it is independent of the message-length. As per our results, our algorithm exhibits strong avalanche effect, as shown in Table II. The results of the both the Chi-Square tests as well as the Monte-Carlo tests, suggests strong randomness as tabulated in Table II.

*5.2 Analysis*

The nonces generated through our algorithm are secure for use in digital signatures due to the following:

a) The resulting nonces from our algorithm are not biased toward the lower range of the DSA domain parameter q, due to the lack of any modulo q operation, which would otherwise have proved fatal [9].

b) Due to the adoption of CSPRNG in generating the round-keys, the output is completely unpredictable and unique each time a message is signed. The inclusion of the message itself in the generation of the nonce adds another layer of randomness and, although the message is known, the processing it goes through inside our algorithm, sufficiently randomize the output. This is suggested from our randomness tests.

c) Since the nonce is generated through our algorithm and the user has no influential control over it, the subliminal channel in DSA has been closed. Although the user can choose the plaintext, the generated nonce cannot be predicted and hence, our algorithm is secure in this respect as well.

d) The algorithm uses non-invertible S-Boxes as well as standard one-way hash functions, and hence, the private-key cannot be recovered from the nonce.

Table I. Constants

| Constant | Value | Used In |
|---|---|---|
| A | bb67 ae85 84ca a73b 2574 2d70 78b8 3b89 25d8 34cc 53da 4798 c720 a648 6e45 a6e2 490b cfd9 5ef1 5dbd a993 0aae 1222 8f87 cc4c f24d a3a1 ec68 d0cd 33a0 1ad9 a383 | Algorithm I |
| P | b9e1 22e6 138c 3ae6 de5e de3b d42d b730 1b6b f553 af7b 09fd 6ebe f33a 9a9f e577 | Algorithm IV |
| Q | 2942 6f30 e589 2ab5 7281 6cce fc58 9935 5f7f 11c3 e24f 3768 a5c7 cb90 | Algorithm VI |
| Z | 243f 6a88 85a3 08d3 1319 8a2e 0370 7344 a409 3822 299f 31d0 082e fa98 ec4e 6c89 4528 21e6 38d0 1377 be54 66cf 34e9 0c6c c0ac 29b7 c97c 50dd 3f84 d5b5 b547 0917 9216 d5d9 8979 fb1b d131 0ba6 98df b5ac 2ffd 72db d01a dfb7 b8e1 afed 6a26 7e96 ba7c 9045 f12c 7f99 24a1 9947 b391 6cf7 0801 f2e2 858e fc16 6369 20d8 7157 4e69 a458 fea3 f493 3d7e 0d95 748f 728e b658 718b cd58 8215 4aee 7b54 a41d c25a 59b5 9c30 d539 2af2 6013 c5d1 b023 2860 85f0 ca41 7918 b8db 38ef 8e79 dcb0 603a 180e 6c9e 0e8b b01e 8a3e d715 77c1 bd31 4b27 78af 2fda 5560 5c60 e655 25f3 aa55 ab94 5748 9862 63e8 1440 55ca 396a 2aab 10b6 b4cc 5c34 1141 e8ce a154 86af 7c72 e993 | (see Section 4.9) |

## 6. CONCLUSION

Digital Signatures form the basis for message authenticity and non-repudiation in modern communication. The entire security of the DSA Algorithm lies in keeping the signer's private-key secret. However, the leaking of bits of the random nonce, seriously compromises this security by enabling an attacker to calculate the private-key from this nonce. Hence, it is extremely crucial that this random nonce be generated in a more secure way that is sufficiently random, unpredictable, and unique for each message. Common pseudorandom sequence generators are insufficient for such purposes, as they can be predictable. Our algorithm improves upon the existing DSA algorithm to enhance its security at the cost of running-time. We believe that this enhanced version of DSA can be successfully used for entity authentication in electronic transactions requiring strict authentication like e-banking [12], e-commerce [23], e-learning [21], and e-governance [22].

Table II. Randomness Analysis

| Test | Results |
|---|---|
| *Avalanche Effect* | When 1-bit in the plaintext was randomly changed, the following were observed: Less than 40% of the bits in the nonce changed = 0.29% of the time 40-50% of the bits in the nonce changed = 65.38% of the time More than 50% of the bits in the nonce changed = 34.33% of the time |
| *Monte-Carlo PI Test* | Maximum Deviation from the correct value of PI: 1.26% Minimum Deviation from the correct value of PI: 0.04% Mean Deviation from the correct value of PI: 0.56% |